\documentclass[twoside,a4paper,12pt]{article}
\usepackage{hyperref}
\pdfoutput=1
\usepackage[utf8]{inputenc}
\usepackage[T1]{fontenc}
\hypersetup{
    pdfauthor={Krešimir Kumerički},
    pdftitle={Feynman Diagrams for Beginners}
}
\usepackage{theorem}
\usepackage{amsmath}
\usepackage{amssymb}
\usepackage{graphicx}
\usepackage{times}
\newcommand{\beq}{\begin{equation}}
\newcommand{\eeq}{\end{equation}}
\newcommand{\pd}{\partial}
\newcommand{\ra}{\rightarrow}
\newcommand{\imp}{\;\Rightarrow\;}

\newcommand{\calM}{\mathcal{M}}
\newcommand{\bra}[1]{\langle #1 |}
\newcommand{\ket}[1]{| #1 \rangle}

\newcommand{\Tr}{\textrm{Tr}}
\newenvironment{example}[1]
{ \vspace*{1em}\noindent \textbf{Example: \emph{#1}}\\[1ex] }
{ $\blacksquare$\vspace*{0.5ex} }

{
\theoremstyle{break}

}
{\theorembodyfont{\upshape}
\theoremstyle{plain}
\newtheorem{exercise}{\itshape Exercise}
}

\if@mathematic
   \def\vec#1{\ensuremath{\mathbf{#1}}}
\else
   \def\vec#1{\ensuremath{\mathchoice{\mbox{\boldmath$\displaystyle#1$}}
                              {\mbox{\boldmath$\textstyle#1$}}
                              {\mbox{\boldmath$\scriptstyle#1$}}
                              {\mbox{\boldmath$\scriptscriptstyle#1$}}}}
\fi

\catcode`\@=11

\def\slash#1{{%
  \setbox\z@\hbox{$\m@th#1$}%
  \setbox\tw@\hbox{$\m@th/$}%
  \dimen@\ht\tw@
  \advance\dimen@-\dp\tw@
  \advance\dimen@-\ht\z@
  \advance\dimen@\dp\z@
  \divide\dimen@\tw@
  \raise-\dimen@\hbox to\wd\z@{\hss\box\tw@\hss}%
  \llap{\box\z@}}}
 
\catcode`\@=12

\usepackage{fancyhdr}
\renewcommand{\subsectionmark}[1]{\relax}

\begin{document}

\title{ \Huge \textbf{Feynman Diagrams for Beginners}\footnote{%
Notes for the exercises at the
\emph{Adriatic School on Particle Physics and Physics Informatics},
11 -- 21 Sep 2001, Split, Croatia}
}

\author{\textbf{Krešimir Kumerički}\footnote{\scriptsize\texttt{kkumer@phy.hr}}\\[2ex]
\small Department of Physics, Faculty of Science, University of Zagreb, Croatia}
\date{}
\maketitle

\begin{abstract}                              
  We give a short introduction to Feynman diagrams, with many exercises.
  Text is targeted at students who had little or no prior exposure to
  quantum field theory. We present condensed description of single-particle 
  Dirac equation,
  free quantum fields and construction of Feynman amplitude using
  Feynman diagrams. As an example, we give a detailed calculation of
  cross-section for annihilation of electron and positron into a muon pair.
  We also show how such calculations are done with the aid of computer.
\end{abstract}      

\tableofcontents

\section{Natural units}

To describe kinematics of some physical system or event we are free to choose
units of measure of the three basic kinematical physical
quantities: \emph{length} \emph{(L)}, \emph{mass} \emph{(M)} and
\emph{time} \emph{(T)}. Equivalently, we may choose any three linearly
independent combinations of these quantities.
The choice of $L$, $T$ and $M$ is usually made (e.g. in SI system of
units) because they are most convenient for description of our immediate
experience. However, elementary particles experience a different world,
one governed by the laws of relativistic quantum mechanics.

Natural units in relativistic quantum mechanics are chosen in such a
way that
fundamental constants of this theory, $c$ and $\hbar$, are both equal to one.
$[c]=LT^{-1}$, $[\hbar]=ML^{-2}T^{-1}$, and to completely fix our system
of units we specify the unit of energy ($ML^2T^{-2}$):
\[
  1\,\mbox{GeV} = 1.6\cdot 10^{-10}\,\mbox{kg m$^2$ s$^{-2}$} \;,
\]
approximately equal to the mass of the proton.
What we do in practice is:
\begin{itemize}
\item we ignore $\hbar$ and $c$ in  formulae and only restore them at the
  end (if at all)
\item we measure \emph{everything} in GeV, GeV$^{-1}$, GeV$^2$, \ldots
\end{itemize}

\begin{example}{Thomson cross section}
Total cross section for scattering of classical electromagnetic radiation by a
free electron (Thomson scattering) is, in natural units,
\begin{equation}
 \sigma_{\rm T}=\frac{8\pi\alpha^2}{3 m_{e}^2} \;.
\end{equation}
To restore $\hbar$ and $c$ we insert them in the above equation with
general powers $\alpha$ and $\beta$, which we determine
by requiring that cross section has the dimension of area ($L^2$):
\begin{equation}
 \sigma_{\rm T}=\frac{8\pi\alpha^2}{3 m_{e}^2}\hbar^{\alpha} c^{\beta}
\end{equation}
\begin{displaymath}
[\sigma] = L^2 = \frac{1}{M^2} (ML^2 T^{-1})^\alpha
(LT^{-1})^\beta
\end{displaymath}
\begin{displaymath}
\imp \alpha = 2\;, \quad \beta = -2  \;,
\end{displaymath}
i.e.
\begin{equation}
 \sigma_{\rm T}=\frac{8\pi\alpha^2}{3 m_{e}^2} \frac{\hbar^2}{c^2}
 = 0.665\cdot 10^{-24} \,\textrm{cm}^2 = 665\,\textrm{mb} \;.
\end{equation}
Linear independence of $\hbar$ and $c$ implies that this can always
be done in a unique way.
\end{example}

Following conversion relations are often useful:

\begin{eqnarray*}
 1\, \textrm{fermi} & = &  5.07\,\textrm{GeV}^{-1} \\ 
 1\, \textrm{GeV}^{-2} & = &  0.389 \,\textrm{mb} \\
 1\, \textrm{GeV}^{-1} & = &  6.582\cdot 10^{-25}\,\textrm{s} \\
 1\, \textrm{kg} & = & 5.61\cdot 10^{26} \,\textrm{GeV} \\
 1\, \textrm{m} & = & 5.07\cdot 10^{15} \,\textrm{GeV}^{-1} \\
 1\, \textrm{s} & = & 1.52\cdot 10^{24} \,\textrm{GeV}^{-1} \\
\end{eqnarray*}

\begin{exercise}
Check these relations.
\end{exercise}

Calculating with GeVs is much more elegant.
Using
$m_e$ =  0.511$\cdot 10^{-3}$ GeV
we get
\begin{equation}
\sigma_{\rm T}=\frac{8\pi\alpha^2}{3 m_{e}^2} = 1709\,\textrm{GeV}^{-2}
= 665 \,\textrm{mb} \;.
\end{equation}
right away.

\begin{exercise}
The decay width of the $\pi^0$ particle is 
\begin{equation}
  \Gamma = \frac{1}{\tau} = 7.7 \,\textrm{eV}.
\end{equation}
Calculate its lifetime $\tau$ in seconds. (By the way,
particle's half-life is equal to $\tau \ln 2$.)
\end{exercise}

\section{Single-particle Dirac equation}

\subsection{The Dirac equation}

Turning the relativistic energy equation
\begin{equation}
   E^2 = \vec{p}^2 + m^2 \;.
\end{equation}
into a differential equation using the usual substitutions
\begin{equation}
     \vec{p} \to -i \nabla \;, \quad E \to i \frac{\pd}{\pd t}\;,
\end{equation}
results in the Klein-Gordon equation:
\begin{equation}
   (\Box + m^2)\psi(x) = 0\;,
\end{equation}
which, interpreted as a single-particle wave equation, has problematic
negative energy solutions. This is due to the negative root in
$E=\pm\sqrt{\vec{p}^2 + m^2}$. Namely, in relativistic \emph{mechanics}
this negative root could be ignored,
but in quantum physics one must keep \emph{all} of the complete set
of solutions to a differential equation.

In order to overcome this problem Dirac tried the ansatz\footnote{
\emph{ansatz}: guess, trial solution (from German \emph{Ansatz}: start, 
beginning, onset, attack)}
\begin{equation}
 (i\beta^{\mu}\pd_{\mu}+m)(i \gamma^{\nu}\pd_{\nu} - m)\psi(x)=0
\label{Dansatz}
\end{equation}
with $\beta^{\mu}$ and $\gamma^{\nu}$ to be determined by requiring
consistency
with the Klein-Gordon equation. This requires $\gamma^{\mu}=\beta^{\mu}$
and 
\begin{equation}
   \gamma^{\mu} \pd_{\mu} \gamma^{\nu} \pd_{\nu} = \pd^{\mu}\pd_{\mu}\;,
\end{equation}
which in turn implies
\begin{displaymath}
 (\gamma^0)^2=1\;,\quad (\gamma^i)^2=-1 \;,
\end{displaymath}
\begin{displaymath}
  \{\gamma^{\mu},\gamma^{\nu}\}\equiv \gamma^{\mu} \gamma^{\nu} +
\gamma^{\nu} \gamma^{\mu} = 0 \quad \mbox{for}\; \mu\neq\nu  \;.
\end{displaymath}
This can be compactly written in form of the \emph{anticommutation relations}
\begin{equation}
   \{\gamma^{\mu},\gamma^{\nu}\} = 2 g^{\mu\nu} \;,\quad
 g^{\mu\nu} = 
 \left(\begin{array}{cccc}
    1 & 0 & 0 & 0 \\
    0 & -1 & 0 & 0 \\
    0 & 0 & -1 & 0 \\
    0 & 0 & 0 & -1 
 \end{array}\right) \;.
\end{equation}
These conditions are obviously impossible to satisfy with $\gamma$'s
being equal to usual numbers, but we can satisfy them by taking
$\gamma$'s equal to (at least) four-by-four matrices.

Now, to satisfy (\ref{Dansatz}) it is enough that one of the two factors
in that equation is zero, and by convention we require this from the second one.
Thus we obtain the \emph{Dirac equation}:
\begin{equation}
 (i \gamma^{\mu} \pd_{\mu} - m)\psi(x) = 0  \;.
\end{equation}
$\psi(x)$ now has four components and is called the \emph{Dirac spinor}.

One of the most frequently used representations for $\gamma$ matrices is
the original Dirac representation
\begin{equation}
\gamma^0 = \left(\begin{array}{cc} 1 & 0 \\
                        0 & -1 \end{array}\right)
\quad
\gamma^i = \left(\begin{array}{cc} 0 & \sigma^i \\
                        -\sigma^i & 0 \end{array}\right) \;,
\end{equation}
where $\sigma^i$ are the Pauli matrices:
\begin{equation}
\sigma^{1}= \left(\begin{array}{rr} 0 & 1 \\
                                    1 & 0 \end{array}\right)
\quad
\sigma^{2}= \left(\begin{array}{rr} 0 & -i \\
                                    i & 0 \end{array}\right)
\quad
\sigma^{3}= \left(\begin{array}{rr} 1 & 0 \\
                                    0 & -1 \end{array}\right)\;.
\end{equation}
This representation is very convenient for the non-relativistic approximation, 
since then the dominant energy terms $(i\gamma^0 \pd_0 - \ldots -m)\psi(0)$ 
turn out to be diagonal.

Two other often used representations are
\begin{itemize}
\item the Weyl (or chiral) representation ---  convenient
    in the ultra-relativistic regime (where $E\gg m$)
\item the Majorana representation --- makes the Dirac equation real; convenient
  for \emph{Majorana fermions} for which antiparticles are equal to particles
\end{itemize}

\noindent
(\emph{Question:} Why can we choose at most one $\gamma$ matrix to be diagonal?)
\\

Properties of the Pauli matrices:
\begin{align}
\sigma^{i^\dagger} &= \sigma^i \\
\sigma^{i*} &= (i\sigma^2)\sigma^i (i\sigma^2) \\
[\sigma^{i}, \sigma^{j}] &= 2 i \epsilon^{ijk}\sigma^{k} \\
\{\sigma^{i}, \sigma^{j}\} &= 2 \delta^{ij} \\
\sigma^{i}\sigma^{j} &= \delta^{ij} + i\epsilon^{ijk}\sigma^{k}
\end{align}
where $\epsilon^{ijk}$ is the totally antisymmetric Levi-Civita tensor
($\epsilon^{123}=\epsilon^{231}=\epsilon^{312}=1$, 
 $\epsilon^{213}=\epsilon^{321}=\epsilon^{132}=-1$, and all other
 components are zero).

\begin{exercise}
 Prove that $(\vec{\sigma}\cdot\vec{a})^2 = \vec{a}^2$ for any 
three-vector \vec{a}.
\end{exercise}

\begin{exercise}
Using properties of the Pauli matrices, prove that $\gamma$
matrices in the Dirac representation satisfy $\{\gamma^i,
\gamma^j\}=2g^{ij}=-2\delta^{ij}$, in accordance with the
anticommutation relations. (Other components of the anticommutation
relations, $(\gamma^0)^2=1$, $\{\gamma^0, \gamma^i\}=0$, are trivial to
prove.)
\end{exercise}

\begin{exercise}
 Show that in the Dirac representation 
$\gamma^0 \gamma^{\mu} \gamma^{0} = \gamma^{\mu^\dagger}$.
\end{exercise}

\begin{exercise}
Determine the Dirac Hamiltonian by writing the Dirac equation in the form
$i\pd\psi/\pd t = H\psi$.
Show that the hermiticity of the Dirac Hamiltonian implies that the relation 
from the previous exercise is valid regardless of the representation.
\end{exercise}

\noindent
The \emph{Feynman slash} notation, $\slash a\equiv a_{\mu}\gamma^{\mu}$,
is often used.

\subsection{The adjoint Dirac equation and the Dirac current}

For constructing the Dirac current we need the equation for $\psi(x)^\dagger$.
By taking the Hermitian adjoint of the Dirac equation we get
\begin{displaymath}
   \psi^\dagger \gamma^{0} (i \stackrel{\leftarrow}{\pd} \;
\!\!\!\!\!\!\!/\; + m)=0\;,
\end{displaymath}
and we define the \emph{adjoint} spinor $\bar{\psi}\equiv
\psi^\dagger \gamma^{0}$ to get the
\emph{adjoint Dirac equation}
\begin{displaymath}
   \bar{\psi}(x) (i \stackrel{\leftarrow}{\pd} \;
\!\!\!\!\!\!\!/\; + m)=0\;.
\end{displaymath}
$\bar{\psi}$ is introduced not only to get aesthetically pleasing equations
but also because it can be shown that, 
unlike $\psi^\dagger$, it transforms covariantly under
the Lorentz transformations.

\begin{exercise}
 Check that the current $j^{\mu} = \bar{\psi} \gamma^{\mu} \psi$ is conserved,
i.e. that it satisfies the continuity relation $\pd_{\mu}j^{\mu}=0$. 
\end{exercise}

Components of this relativistic four-current are $j^{\mu}=(\rho, \vec{j})$. 
Note that $\rho = j^0 = \bar{\psi} \gamma^{0} \psi = \psi^\dagger \psi
 > 0$, i.e. that probability is positive definite, as it must be.

\subsection{Free-particle solutions of the Dirac equation}

Since we are preparing ourselves 
for the perturbation theory calculations, we need to consider
only free-particle solutions.
For solutions in various potentials, see the literature.

The fact that Dirac spinors satisfy the Klein-Gordon equation suggests
the ansatz
\begin{equation}
            \psi(x) = u(\vec{p}) e^{-ipx}\;,
\end{equation}
which after inclusion in the Dirac equation gives the \emph{momentum space
Dirac equation}
\begin{equation}
   (\slash p -m)u(\vec{p})=0\;.
\end{equation}
This has two positive-energy solutions
\begin{equation}
\renewcommand{\arraystretch}{1.8}
 u(\vec{p},\sigma) = N 
\left( \begin{array}{c} \chi^{(\sigma)} \\
{\displaystyle \frac{\vec{\sigma}\cdot\vec{p}}{E+m}}\,\chi^{(\sigma)}
\end{array} \right)\;, \quad \sigma=1,2\;,
\label{u}
\end{equation}
where
\begin{equation}
 \chi^{(1)}=\left( \begin{array}{c} 1 \\ 0 \end{array} \right)\;,\quad
 \chi^{(2)}=\left( \begin{array}{c} 0 \\ 1 \end{array} \right)  \;,
\end{equation}
and two negative-energy solutions which are then interpreted as
positive-energy \emph{antiparticle} solutions
\begin{equation}
\renewcommand{\arraystretch}{1.8}
 v(\vec{p},\sigma) = -N 
\left( \begin{array}{c} 
{\displaystyle \frac{\vec{\sigma}\cdot\vec{p}}{E+m}}\,(i\sigma^2)\chi^{(\sigma)} \\
(i\sigma^2)\chi^{(\sigma)} 
\end{array} \right)\;, \quad \sigma=1,2, \qquad E>0 \;.
\label{v}
\end{equation}
$N$ is the normalization constant to be determined later.
Spinors above agree with those of \cite{Griffiths:2008zz}.
The momentum-space Dirac equation for antiparticle solutions is
\begin{equation}
         (\slash p +m)v(\vec{p},\sigma) = 0 \;.
\end{equation}

It can be shown that the two solutions, one with $\sigma=1$ and
another with $\sigma=2$, correspond to the two spin states of the
spin-1/2 particle.

\begin{exercise}
 Determine momentum-space Dirac equations for $\bar{u}(\vec{p},\sigma)$
and $\bar{v}(\vec{p},\sigma)$.
\end{exercise}

\subsubsection{Normalization}

 In non-relativistic single-particle quantum mechanics normalization of
a wavefunction is straightforward. Probability that the
particle is somewhere in space is equal to one, 
and this translates into the normalization
condition $\int \psi^* \psi\, dV = 1$.
On the other hand, we will eventually use spinors (\ref{u}) and
(\ref{v}) in many-particle quantum field theory so their
normalization is not unique. We will choose normalization convention
where we have $2E$ particles in the unit volume:
\begin{equation}
  \int\limits_{\mbox{\scriptsize unit volume}}\rho\, dV =
  \int\limits_{\mbox{\scriptsize unit volume}} \psi^\dagger \psi \,dV = 2E
\end{equation}

\noindent
This choice is relativistically covariant because the Lorentz contraction of
the volume element is compensated by the energy change.
There are other normalization conventions with other advantages.

\begin{exercise}
Determine the normalization constant N conforming to this choice.
\end{exercise}

\subsubsection{Completeness}

\begin{exercise}
 Using the explicit expressions (\ref{u}) and (\ref{v}) show that
\begin{eqnarray}
\sum_{\sigma=1,2} u(\vec{p},\sigma) \bar{u}(\vec{p},\sigma) &=&\slash p + m\;,
\label{ucomplete} \\
\sum_{\sigma=1,2} v(\vec{p},\sigma) \bar{v}(\vec{p},\sigma) &=&\slash p - m \;.
\label{vcomplete}
\end{eqnarray}
\end{exercise}
  These relations are often needed in calculations of Feynman diagrams with
unpolarized fermions. See later sections.

\subsubsection{Parity and bilinear covariants}

The parity transformation:
\begin{itemize}
\item  $P: \vec{x} \to -\vec{x}$, $t \to t$
\item  $P: \psi \to \gamma^0 \psi$
\end{itemize}

\begin{exercise}
Check that the current $j^{\mu} = \bar{\psi} \gamma^{\mu} \psi$ transforms
as a vector under parity i.e. that $j^0 \to j^0$ and $\vec{j}\to -\vec{j}$.
\end{exercise}

Any fermion current will be of the form $\bar{\psi}\Gamma\psi$, where
$\Gamma$ is some four-by-four matrix. For construction of interaction
Lagrangian we want to use only those currents that have definite
Lorentz transformation properties. To this end we first define
two new matrices:

\begin{equation}
\gamma^5 \equiv i\gamma^0\gamma^1\gamma^2\gamma^3\stackrel{\textrm{Dirac rep.}}
{=}
\left(\begin{array}{rr} 0 & 1 \\ 1 & 0 \end{array}\right)\;, \quad
\{\gamma^5,\gamma^{\mu}\}=0\;,
\end{equation}

\begin{equation}
\sigma^{\mu\nu}\equiv\frac{i}{2}[\gamma^{\mu}, \gamma^{\nu}]\;, \quad
\sigma^{\mu\nu}=-\sigma^{\nu\mu}\;.
\end{equation}

Now $\bar{\psi}\Gamma\psi$ will transform covariantly if $\Gamma$ is
one of the matrices given in the following table. Transformation
properties of $\bar{\psi}\Gamma\psi$, the number of different
$\gamma$ matrices in $\Gamma$, and the number of components of $\Gamma$
are also displayed.

\vspace*{3ex}
\centerline{
\begin{tabular}{cccc}
\hline $\Gamma$ & transforms as & \# of $\gamma$'s & \# 
of components\\\hline 
$1$      &   scalar      &        0        &     1         \\
$\gamma^{\mu}$ & vector      &        1        &     4     \\
$\sigma^{\mu\nu}$ & tensor     &        2        &     6    \\
$\gamma^5\gamma^{\mu}$ & axial vector  &  3      &     4     \\
$\gamma^5$ & pseudoscalar   &  4      &                 1     \\
\hline 
\end{tabular}
}
\vspace*{4ex}

\noindent
This exhausts all possibilities.
The total number of components is 16, meaning that the set $\{1, \gamma^{\mu}, 
\sigma^{\mu\nu}, \gamma^5\gamma^{\mu}, \gamma^5\}$ makes a complete basis
for any four-by-four matrix. Such $\bar{\psi}\Gamma\psi$ currents are
called \emph{bilinear covariants}.

\section{Free quantum fields}

Single-particle Dirac equation is (a) not exactly right even for single-particle
systems such as the H-atom, and (b) unable to treat many-particle processes
such as the $\beta$-decay $n\to p\, e^- \bar{\nu}$. We have to upgrade 
to quantum field theory.

Any Dirac field is some superposition of the complete set
\begin{displaymath}
u(\vec{p},\sigma)e^{-ipx}\;, \quad v(\vec{p},\sigma)e^{ipx} \;,
\quad \sigma=1,2, \quad \vec{p}\in\mathbb{R}^3
\end{displaymath}
and we can write it as
\begin{equation}
\psi(x)=\sum_\sigma \int \frac{d^3 p}{\sqrt{(2\pi)^3 2 E}}
\left[ u(\vec{p},\sigma) a(\vec{p},
\sigma) e^{-i p x} + v(\vec{p},\sigma)
a^{c\dagger}(\vec{p},\sigma) e^{i p x}\right] \;.
\label{Diracfield}
\end{equation}
Here $1/\sqrt{(2\pi)^3 2 E}$ is a normalization factor (there are many
different conventions), and $a(\vec{p},\sigma)$ and 
$a^{c\dagger}(\vec{p},\sigma)$ are expansion coefficients.
To make this a \emph{quantum Dirac field} we promote these
coefficients to the rank of operators by imposing the \emph{anticommutation}
 relations
\begin{equation}
           \{a(\vec{p},\sigma), a^{\dagger}(\vec{p}',\sigma')\}
= \delta_{\sigma\sigma'}\delta^{3}(\vec{p}-\vec{p}'),
\end{equation}
and similarly for $a^{c}(\vec{p},\sigma)$.
(For bosonic fields we would have a \emph{commutation} relations instead.)
This is similar to the promotion of position and momentum to the rank
of operators by the $[x_i,p_j]=i\hbar\delta_{ij}$ commutation relations, which
is why is this transition from the single-particle quantum theory to the quantum
field theory sometimes called \emph{second quantization}. 

Operator $a^\dagger$, when operating on vacuum state  $\ket{0}$, creates
one-particle state $\ket{\vec{p},\sigma}$
\begin{equation}
  a^{\dagger}(\vec{p},\sigma)\ket{0} = \ket{\vec{p},\sigma} \;,
\end{equation}
and  this is the reason that it is named a \emph{creation} operator.
Similarly, $a$ is an \emph{annihilation} operator
\begin{equation}
  a(\vec{p},\sigma)\ket{\vec{p},\sigma} = \ket{0} \;,
\end{equation}
and $a^{c\dagger}$ and $a^{c}$ are creation and annihilation operators
for antiparticle states ($c$ in $a^{c}$ stands for ``conjugated'').

Processes in particle physics are mostly calculated in the framework
of the theory of such fields --- \emph{quantum field theory}. This
theory can be described at various levels of rigor but in any case
is complicated enough to be beyond
the scope of these notes.

  However, predictions of quantum field theory pertaining to the elementary 
particle interactions can often be calculated using a relatively
simple ``recipe'' --- \emph{Feynman diagrams}.

 Before we turn to describing the method of Feynman diagrams, 
let us just specify other
quantum fields that take part in the elementary particle physics interactions.
All these are \emph{free} fields, and interactions are treated as their
perturbations. Each particle type (electron, photon, Higgs boson, ...)
has its own quantum field.

\subsection{Spin 0: scalar field}

E.g. Higgs boson, pions, ...

\begin{equation}
\phi(x)= \int \frac{d^3 p}{\sqrt{(2\pi)^3 2 E}}
\left[  a(\vec{p}) e^{-i p x} + 
a^{c\dagger}(\vec{p}) e^{i p x}\right] 
\label{scalarfield}
\end{equation}

\subsection{Spin 1/2: the Dirac field}

E.g. quarks, leptons

\vspace{1ex}

We have already specified the Dirac spin-1/2 field. There are other types: Weyl
and Majorana spin-1/2 fields but they are beyond our scope.

\subsection{Spin 1: vector field}

Either
\begin{itemize}
\item  massive (e.g. W,Z weak bosons) or
\item massless (e.g. photon)
\end{itemize}

\begin{equation}
A^{\mu}(x)=\sum_\lambda \int \frac{d^3 p}{\sqrt{(2\pi)^3 2 E}}
\left[ \epsilon^{\mu}(\vec{p},\lambda) a(\vec{p},
\lambda) e^{-i p x} + \epsilon^{\mu\ast}(\vec{p},\lambda)
a^{\dagger}(\vec{p},\lambda) e^{i p x}\right] 
\label{vectorfield}
\end{equation}

$\epsilon^{\mu}(\vec{p},\lambda)$ is a polarization vector. For
massive particles it obeys
\begin{equation}
     p_{\mu}\epsilon^{\mu}(\vec{p},\lambda)=0
\end{equation}
automatically, whereas in the massless case this condition can be
imposed thanks to gauge invariance (Lorentz gauge condition). 
This means that there are only three independent polarizations
of a massive vector particle:
$\lambda=1,2,3$ or $\lambda=+,-,0$.
In massless case gauge symmetry can be further exploited to 
eliminate one more polarization state leaving us with only two:
$\lambda=1,2$ or $\lambda=+,-$.

Normalization of polarization vectors is such that
\begin{equation}
  \epsilon^*(\vec{p},\lambda)\cdot\epsilon(\vec{p},\lambda) = -1\;.
\end{equation}

E.g. for a massive particle moving along the
$z$-axis ($p=(E,0,0,|\vec{p}|)$) we can
take 
\begin{equation}
 \epsilon(\vec{p},\pm) = \mp \frac{1}{\sqrt{2}} 
\left( \begin{array}{c} 0 \\ 1 \\ \pm i \\ 0 \end{array} \right)\;, \quad
 \epsilon(\vec{p},0) =  \frac{1}{m} 
\left( \begin{array}{c} |\vec{p}| \\ 0 \\ 0 \\ E \end{array} \right) 
\end{equation}

\begin{exercise}
 Calculate
\begin{displaymath}
      \sum_{\lambda}\epsilon^{\mu\ast}(\vec{p},\lambda)
\epsilon^{\nu}(\vec{p},\lambda)
\end{displaymath}
Hint: Write it in the most general form $(Ag^{\mu\nu}+Bp^{\mu}p^{\nu})$
and then determine $A$ and $B$.
\end{exercise}

The obtained result obviously cannot be simply extrapolated to the massless
case via the
limit $m\to 0$. Gauge symmetry makes massless polarization sum
somewhat more complicated but for the purpose of the
simple Feynman diagram calculations it is
permissible to use just the following relation

\begin{displaymath}
      \sum_{\lambda}\epsilon^{\mu\ast}(\vec{p},\lambda)
\epsilon^{\nu}(\vec{p},\lambda) = - g^{\mu\nu} \;.
\end{displaymath}

\section{Golden rules for decays and scatterings}
\label{goldenrules}

Principal experimental observables of particle physics are
\begin{itemize}
\item scattering cross section $\sigma(1+2\to 1'+2'+\cdots +n')$
\item decay width $\Gamma(1\to 1'+2'+\cdots +n')$ 
\end{itemize}

On the other hand, theory is defined in terms of Lagrangian density
of quantum fields, e.g.
\begin{displaymath}
\mathcal{L}=\frac{1}{2}\pd_{\mu}\phi\pd^{\mu}\phi - \frac{1}{2}m^2\phi^2
-\frac{g}{4!}\phi^4\;.
\end{displaymath}

How to calculate $\sigma$'s and $\Gamma$'s from $\mathcal{L}$?

To calculate rate of transition from the
state $\ket{\alpha}$ to the state $\ket{\beta}$ in the presence of the 
interaction potential $V_I$ in non-relativistic quantum theory
we have the Fermi's Golden Rule
\begin{equation}
{ \alpha \to \beta \choose \mbox{transition rate} } = \frac{2\pi}{\hbar}
|\bra{\beta}V_I \ket{\alpha}|^2 \times { \mbox{density of final} \choose
 \mbox{quantum states}} \;.
\end{equation}
This is in the lowest order perturbation theory. For higher orders we have
terms with products of more interaction potential matrix elements
$\bra{}V_I\ket{}$.

In quantum field theory there is a counterpart to these matrix elements
--- the \emph{S-matrix}: 
\begin{equation}
 \bra{\beta}V_I \ket{\alpha} + (\mbox{\scriptsize higher-order terms})
\quad\longrightarrow\quad \bra{\beta} S \ket{\alpha}\;.
\end{equation}

On one side, $S$-matrix elements can be perturbatively calculated 
(knowing the interaction Lagrangian/Hamiltonian) with the help of the
\emph{Dyson series}
\begin{equation}
S=1-i\int d^4x_1\, \mathcal{H}(x_1) + \frac{(-i)^2}{2!} \int d^4x_1\,d^4x_2 \,
T\{\mathcal{H}(x_1)\mathcal{H}(x_2)\} + \cdots  \;,
\label{Dyson}
\end{equation}
and on another, we have ``golden rules'' that associate these
matrix elements with cross-sections and decay widths.

It is convenient to express these golden rules in terms of the 
\emph{Feynman invariant amplitude}  $\mathcal{M}$
which is obtained by stripping some kinematical
factors off the $S$-matrix:
\begin{equation}
\bra{\beta}S\ket{\alpha}=\delta_{\beta\alpha} - i (2\pi)^4 \delta^{4}(p_\beta -
p_\alpha) {\cal M}_{\beta\alpha} \prod_{i=\alpha,\beta}\frac{1}{
\sqrt{(2\pi)^3\, 2E_{i}}} \;.
\end{equation}
Now we have two rules:
\begin{itemize}
\item  
Partial decay rate of $1\ra 1'+2'+\cdots +n'$
\begin{equation}
d\Gamma = \frac{1}{2E_1}\overline{|{\cal M}_{\beta\alpha}|^2}\:
 (2\pi)^4 \delta^4(p_1-p'_1-\cdots-p'_n)\prod_{i=1}^{n} \frac{d^3 p'_i}{(2\pi)^3 
\,2E'_{i}} \;,
\label{goldenGamma}
\end{equation}
\item
Differential cross section for a scattering $1+2\ra 1'+2'+\cdots+n'$
\begin{equation}
d\sigma = \frac{1}{u_\alpha}\frac{1}{2 E_1}\frac{1}{2 E_2}
 \overline{|{\cal M}_{\beta\alpha}|^2}\:
 (2\pi)^4 \delta^4(p_1+p_2-p'_1-\cdots-p'_n)
   \prod_{i=1}^{n} \frac{d^3 p'_i}{(2\pi)^3
\,2E'_{i}} \;,
\label{goldensigma}
\end{equation}
\end{itemize}
where $u_\alpha$ is the relative velocity of particles 1 and 2:
\begin{equation}
u_\alpha =\frac{\sqrt{(p_1\cdot p_2)^2 -m_{1}^2 m_{2}^2}}{E_1 E_2} \;,
\end{equation}
and $\overline{|{\cal M}|^2}$ is the Feynman invariant amplitude averaged over
unmeasured particle spins (see Section \ref{polaverage}). 
The dimension of ${\cal M}$, in units of energy, is
\begin{itemize}
\item for decays $[{\cal M}]=3-n$
\item for scattering of two particles $[{\cal M}]=2-n$
\end{itemize}
where $n$ is the number of produced particles.

\vspace*{1ex}

\noindent
So calculation of some observable quantity consists of two stages:
\begin{enumerate}
\item Determination of $\overline{|{\cal M}|^2}$. For this we use the
method of Feynman diagrams to be introduced in the next section.
\item Integration over the Lorentz invariant phase space
\begin{displaymath}
d\mbox{Lips}=
 (2\pi)^4 \delta^4(p_1+p_2-p'_1-\cdots-p'_n)
   \prod_{i=1}^{n} \frac{d^3 p'_i}{(2\pi)^3
\,2E'_{i}} \;.
\end{displaymath}
\end{enumerate}

\section{Feynman diagrams}

\begin{example}{$\phi^{4}$-theory}

\begin{displaymath}
\mathcal{L}=\frac{1}{2}\pd_{\mu}\phi\pd^{\mu}\phi - \frac{1}{2}m^2\phi^2
-\frac{g}{4!}\phi^4
\end{displaymath}

\begin{itemize}
\item Free (kinetic) Lagrangian (terms with exactly two fields) determines 
particles of the theory and their propagators.  
Here we have just one scalar field:

\centerline{\includegraphics[scale=1]{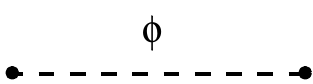}}

\item Interaction Lagrangian (terms with three or more fields) determines 
possible vertices. Here, again, there is just one vertex:

\centerline{\includegraphics[scale=1]{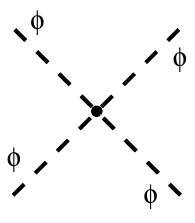}}

\end{itemize}

We construct all possible diagrams with fixed outer particles. E.g.
for scattering of two scalar particles in this theory we would have

${\cal M}(1+2\to 3+4)$ =
\parbox{10em}{\includegraphics[scale=1]{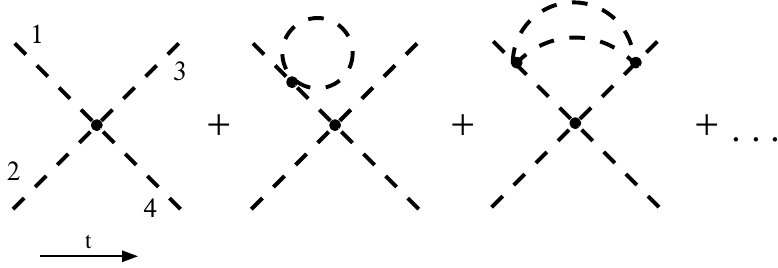}}

In these diagrams time flows from left to right. Some 
people draw Feynman diagrams with time flowing up, which is more in accordance
with the way we usually draw space-time in relativity physics.

Since each vertex corresponds to one interaction Lagrangian (Hamiltonian)
term in (\ref{Dyson}), diagrams with loops correspond to higher
orders of perturbation theory. Here we will work only to the lowest order,
so we will use \emph{tree diagrams} only.

To actually write down the Feynman amplitude $\mathcal{M}$, we have a set of
\emph{Feynman rules} that associate factors with elements of the Feynman
diagram.
In particular, to get $-i\mathcal{M}$ we construct the Feynman rules in
the following way:
\begin{itemize}
\item the vertex factor
is just the $i$ times the interaction term in the (momentum space)
Lagrangian with all fields removed:
\begin{equation}
i\mathcal{L}_{\rm I}=-i \frac{g}{4!}\phi^4 \quad
\stackrel{\mbox{\scriptsize removing fields}}{\imp}\quad
\parbox{3em}{\includegraphics[scale=1.0]{phi_vert}} =
-i \frac{g}{4!}
\end{equation}

\item the propagator is $i$ times the inverse of the kinetic operator (defined
by the free equation of motion) in
the momentum space:
\begin{equation}
\mathcal{L}_{\rm free} \stackrel{\mbox{\scriptsize Euler-Lagrange eq.}}{
\longrightarrow} (\pd_{\mu}\pd^{\mu}+m^2)\phi=0 \qquad \mbox{(Klein-Gordon eq.)}
\end{equation}
Going to the momentum space using the substitution $\pd^{\mu} \to -i p^{\mu}$
and then taking the inverse gives:
\begin{equation}
(p^2-m^2)\phi=0 \quad\imp\quad
\parbox{8em}{\includegraphics[scale=1.0]{phi_prop}} =
 \frac{i}{p^2-m^2}
\end{equation}
(Actually, the correct Feynman propagator is $i/(p^2-m^2+i\epsilon)$, but
for our purposes we can ignore the infinitesimal $i\epsilon$ term.)

\item External lines are represented by the appropriate polarization
vector or spinor (the one that stands by the appropriate creation or
annihilation operator in the fields (\ref{Diracfield}), (\ref{scalarfield}),
(\ref{vectorfield}) and their conjugates):

\vspace*{2ex}
\begin{tabular}{lc}
\hline particle & Feynman rule  \\ \hline
ingoing fermion  &  $u$  \\
outgoing fermion  &  $\bar{u}$  \\
ingoing antifermion  &  $\bar{v}$  \\
outgoing antifermion  &  $v$  \\
ingoing photon  &  $\epsilon^{\mu}$\\
outgoing photon  &  $\epsilon^{\mu\ast}$\\
ingoing scalar  &  1 \\
outgoing scalar  &  1 \\ \hline
\end{tabular}
\vspace*{5ex}

\end{itemize}

  So the tree-level contribution to the scalar-scalar scattering amplitude
in this $\phi^4$ theory would be just
\begin{equation}
    -i\mathcal{M} = -i \frac{g}{4!} \;.
\end{equation}

\end{example}

\begin{exercise}
Determine the Feynman rules for the electron propagator and for the only
vertex of quantum electrodynamics (QED):
\begin{equation}
\mathcal{L}=\bar{\psi}(i\slash \pd + e \slash A -m)\psi -
\frac{1}{4}F_{\mu\nu}F^{\mu\nu} \qquad F^{\mu\nu}=\pd^{\mu}A^{\nu}-
\pd^{\nu}A^{\mu} \;.
\end{equation}
\end{exercise}

Note that also
\begin{equation}
\parbox{8em}{\includegraphics[scale=1.0]{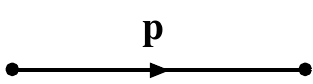}} =
\frac{i\sum_{\sigma}u(\vec{p},\sigma)\bar{u}(\vec{p},\sigma)}
{p^2-m^2} \;,
\end{equation}
i.e. the electron propagator is just the scalar propagator multiplied by the
polarization sum.  It is nice that this generalizes to propagators of all
particles. This is very helpful since inverting the photon kinetic operator is
non-trivial due to gauge symmetry complications. Hence, propagators of vector
particles are

\vspace*{1ex}
\begin{equation}
\mbox{massive:}\qquad\quad
\parbox{8em}{\includegraphics[scale=1.0]{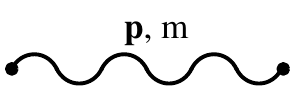}}  = 
\frac{-i\left(g^{\mu\nu}-{\displaystyle 
\frac{p^{\mu}p^{\nu}}{m^2}}\right)}{p^2-m^2}\;, \\
\end{equation}
\vspace*{1ex}
\begin{equation}
\mbox{massless:}\qquad\quad
\parbox{8em}{\includegraphics[scale=1.0]{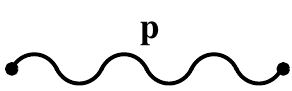}}  = 
\frac{-ig^{\mu\nu}}{p^2}\;.
\end{equation}
\vspace*{1ex}

This is in principle \emph{almost} all we need to know 
to be able to calculate the
Feynman amplitude of any given process.
Note that propagators and external-line polarization vectors are determined
only by the particle type (its spin and mass) so that
the corresponding rules above are not
restricted only to the $\phi^4$ theory and QED,
but will apply to all theories of
scalars, spin-1 vector bosons and Dirac fermions (such as the standard
model). The only additional information we need are the vertex factors.

\emph{``Almost''} in the preceding paragraph alludes to the fact
that in general Feynman diagram calculation there are several additional
subtleties:
\begin{itemize}
\item In loop diagrams some internal momenta are undetermined and we
 have to integrate over those. Also, there is an additional factor (-1)
 for each closed fermion loop. Since we will consider tree-level diagrams only,
  we can ignore this.
\item There are some combinatoric numerical factors when identical 
 fields come into a single vertex.
\item Sometimes there is a relative (-) sign between diagrams.
\item There is a symmetry factor if there are identical particles
   in the final state.
\end{itemize}

For explanation of these, reader is advised to look in some quantum field
theory textbook.

\section{Example: \boldmath $e^+ e^- \to \mu^+ \mu^-$\unboldmath\hspace*{0.5em} in QED}

There is only one contributing tree-level diagram:

\vspace*{1ex}
\centerline{\includegraphics[scale=1.0]{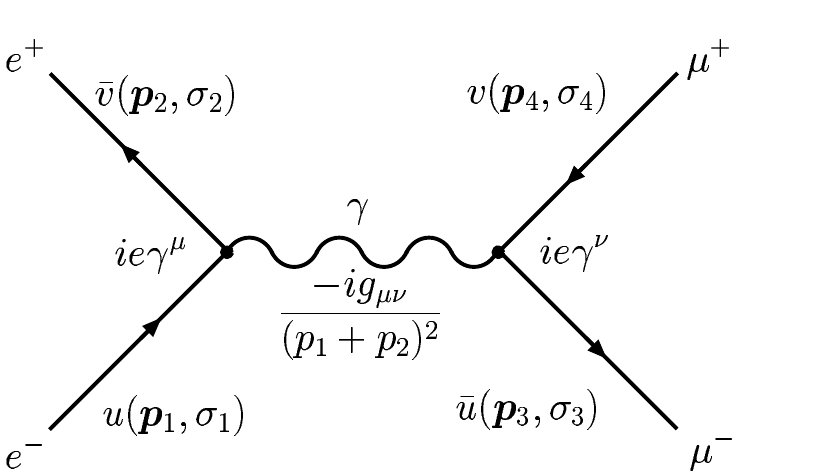}} 
\vspace*{1ex}

We write down the amplitude using the Feynman rules of QED
and following fermion
lines backwards. Order of lines themselves is unimportant.

\begin{equation}
-i\calM = \left[\bar{u}(\vec{p}_3,\sigma_3) (ie \gamma^{\nu})  
v(\vec{p}_4,\sigma_4) \right] \frac{-i g_{\mu\nu}}{(p_1 + p_2)^2}
\left[ 
\bar{v}(\vec{p}_2,\sigma_2) (ie \gamma^{\mu}) u(\vec{p}_1,\sigma_1)\right]\;,
\end{equation}

or, introducing abbreviation $u_1 \equiv u(\vec{p}_1,\sigma_1)$,

\begin{equation}
\calM = \frac{e^2}{(p_1 + p_2)^2} [\bar{u}_3 \gamma_{\mu} v_4 ]
[\bar{v}_2 \gamma^{\mu} u_1] \;.
\end{equation}

\begin{exercise}
Draw Feynman diagram(s) and write down the amplitude for Compton scattering
$\gamma e^{-} \to \gamma e^{-}$.
\end{exercise}

\subsection{Summing over polarizations}
\label{polaverage}

If we knew momenta and polarizations of all external particles, we could
calculate $\calM$ explicitly. However, experiments are often done with
unpolarized particles so we have to sum over the polarizations (spins) 
of the final
particles and average over the polarizations (spins) of the initial ones:

\begin{equation}
|\calM |^2 \to \overline{|\calM |^2}=\underbrace{\frac{1}{2}\; \frac{1}{2}\;
\sum_{\sigma_1 \sigma_2}}_{\mbox{\scriptsize avg. over initial pol.}}
\overbrace{\sum_{\sigma_3 \sigma_4}}^{\mbox{\scriptsize sum over final pol.}}
|\calM |^2\;.
\end{equation}

Factors $1/2$ are due to the fact that each initial fermion has two
polarization (spin) states.

\noindent
(\emph{Question:} Why we sum probabilities and not amplitudes?)

\vspace*{1ex}

In the calculation of
$|\calM|^2 =\calM^* \calM$, the following identity is needed

\begin{equation}
[\bar{u} \gamma^{\mu} v]^* = [u^{\dagger}\gamma^{0} \gamma^{\mu} v]^{\dagger}
= v^{\dagger}\gamma^{\mu\dagger}\gamma^{0} u = [\bar{v}\gamma^{\mu}u]\;.
\end{equation}

Thus,

\begin{equation}
\overline{|\calM |^2} = \frac{e^4}{4(p_1 + p_2)^4} \sum_{\sigma_{1,2,3,4}}
[\bar{v}_4 \gamma_{\mu} u_3 ]
[\bar{u}_1 \gamma^{\mu} v_2]
[\bar{u}_3 \gamma_{\nu} v_4 ]
[\bar{v}_2 \gamma^{\nu} u_1]\;.
\end{equation}

\subsection{Casimir trick}

Sums over polarizations are easily performed using the following trick.
First we write $\sum [\bar{u}_1 \gamma^{\mu} v_2][\bar{v}_2 \gamma^{\nu} u_1]$
with explicit spinor indices $\alpha,\beta,\gamma,\delta=1,2,3,4$:
\begin{equation}
\sum_{\sigma_1 \sigma_2} 
\bar{u}_{1\alpha} \gamma^{\mu}_{\alpha\beta} v_{2\beta}\;
\bar{v}_{2\gamma} \gamma^{\nu}_{\gamma\delta} u_{1\delta} \;.
\label{casimirsum}
\end{equation}
We can now move $u_{1\delta}$ to the front ($u_{1\delta}$ is just a number,
element of $u_1$ vector, so it commutes with everything), and then
use the completeness relations (\ref{ucomplete}) and (\ref{vcomplete}),
\begin{eqnarray*}
\sum_{\sigma_1} u_{1\delta} \,\bar{u}_{1\alpha} &=&(\slash p_1 + m_1)_{
\delta\alpha}\;,
 \\
\sum_{\sigma_2} v_{2\beta} \,\bar{v}_{2\gamma} &=&(\slash p_2 - m_2)_{
\beta\gamma} \;,
\end{eqnarray*}
which turn sum (\ref{casimirsum}) into
\begin{equation}
 (\slash p_1 + m_1)_{\delta\alpha}\, \gamma^{\mu}_{\alpha\beta}\,
(\slash p_2 - m_2)_{\beta\gamma}\, \gamma^{\nu}_{\gamma\delta}
= \mbox{Tr}[(\slash p_1 +m_1)\gamma^{\mu} (\slash p_2 -m_2)\gamma^{\nu}]\;.
\end{equation}

This means that 

\begin{equation}
\overline{|\calM |^2} = \frac{e^4}{4(p_1 + p_2)^4}
\Tr[(\slash p_1 +m_1)\gamma^{\mu} (\slash p_2 -m_2)\gamma^{\nu}]\,
\Tr[(\slash p_4 -m_4)\gamma_{\mu} (\slash p_3 +m_3)\gamma_{\nu}]\;.
\end{equation}

Thus we got rid off all the spinors and we are left only with traces of
$\gamma$ matrices.
These can be evaluated using the relations from the following section.

\subsection{Traces and contraction identities of \boldmath$\gamma\,$\unboldmath
 matrices}

All are consequence of the anticommutation relations
$\{\gamma^{\mu},\gamma^{\nu}\} = 2 g^{\mu\nu}$, $\{\gamma^{\mu},\gamma^{5}\} =
0$, $(\gamma^5)^2=1$, and of nothing else!

\subsubsection{Trace identities}

\begin{enumerate}

\item Trace of an odd number of $\gamma$'s vanishes:
\begin{eqnarray*}
\Tr(\gamma^{\mu_1}\gamma^{\mu_2}\cdots\gamma^{\mu_{2n+1}}) & = & 
\Tr(\gamma^{\mu_1}\gamma^{\mu_2}\cdots\gamma^{\mu_{2n+1}}
\overbrace{\gamma^5\gamma^5}^{1}) \\
\mbox{\scriptsize (moving $\gamma^5$ over each $\gamma^{\mu_i}$)}& = &
-\Tr(\gamma^5\gamma^{\mu_1}\gamma^{\mu_2}\cdots\gamma^{\mu_{2n+1}}
\gamma^5) \\
\mbox{\scriptsize (cyclic property of trace)}& = & 
-\Tr(\gamma^{\mu_1}\gamma^{\mu_2}\cdots\gamma^{\mu_{2n+1}}
\gamma^5\gamma^5) \\
& = &
- \Tr(\gamma^{\mu_1}\gamma^{\mu_2}\cdots\gamma^{\mu_{2n+1}}) \\
& = & 0
\end{eqnarray*}

\item Tr 1 = 4

\item 
\begin{displaymath}
 \Tr \gamma^{\mu} \gamma^{\nu} = \Tr (2g^{\mu\nu}-\gamma^{\nu}\gamma^{\mu})
\stackrel{(2.)}{=} 8 g^{\mu\nu} -\Tr \gamma^{\nu} \gamma^{\mu} =
                  8 g^{\mu\nu} -\Tr \gamma^{\mu} \gamma^{\nu}
\end{displaymath}
\begin{displaymath}
\imp 2\Tr \gamma^{\mu} \gamma^{\nu}=8 g^{\mu\nu} \imp
\Tr \gamma^{\mu} \gamma^{\nu} = 4 g^{\mu\nu}
\end{displaymath}
This also implies:
\begin{displaymath}
  \Tr \slash a \slash b = 4 a\cdot b
\end{displaymath}

\item
\begin{exercise}
 Calculate $\Tr(\gamma^{\mu} \gamma^{\nu} \gamma^{\rho} \gamma^{\sigma})$.
Hint: Move $\gamma^{\sigma}$ all the way to the left, using the anticommutation
relations. Then use 3.
\end{exercise}

\noindent
\emph{Homework:} Prove that 
$\Tr(\gamma^{\mu_1}\gamma^{\mu_2}\cdots\gamma^{\mu_{2n}})$ 
has $(2n-1)!!$ terms.

\item
$\Tr(\gamma^{5}\gamma^{\mu_1}\gamma^{\mu_2}\cdots\gamma^{\mu_{2n+1}}) = 0$.
This follows from 1. and from the fact that $\gamma^{5}$ consists of
even number of $\gamma$'s.

\item $\Tr \gamma^{5} = \Tr(\gamma^{0} \gamma^{0} \gamma^{5})
= -\Tr(\gamma^{0} \gamma^{5} \gamma^{0} ) = - \Tr \gamma^{5} = 0$

\item $\Tr(\gamma^{5}\gamma^{\mu} \gamma^{\nu})=0$. (Same trick as above,
with $\gamma^{\alpha}\neq \mu,\nu$ instead of $\gamma^{0}$.)

\item
$\Tr(\gamma^{5}\gamma^{\mu} \gamma^{\nu} \gamma^{\rho} \gamma^{\sigma})
= - 4 i \epsilon^{\mu\nu\rho\sigma}$, with $\epsilon^{0123}=1$.
Careful: convention with $\epsilon^{0123}=-1$ is also in use.
\end{enumerate}

\subsubsection{Contraction identities}

\begin{enumerate}

\item
\begin{displaymath}
\gamma^{\mu} \gamma_{\mu} = \frac{1}{2} g_{\mu\nu}
\underbrace{(\gamma^{\mu} \gamma^{\nu} + \gamma^{\nu} \gamma^{\mu})}_{
 2 g^{\mu\nu}}=g_{\mu\nu}g^{\mu\nu}=4
\end{displaymath}

\item 
\begin{displaymath}
\gamma^{\mu} \!\!\!\!\! 
\underbrace{\gamma^{\alpha} \gamma_{\mu}}_{-\gamma_{\mu}
\gamma^{\alpha} + 2g_{\mu}^{\alpha}} = -4 \gamma^{\alpha} +2\gamma^{\alpha}
=-2\gamma^{\alpha}
\end{displaymath}

\item
\begin{exercise}
  Contract $\gamma^{\mu} \gamma^{\alpha}\gamma^{\beta}\gamma_{\mu}$.
\end{exercise}

\item $\gamma^{\mu}\gamma^{\alpha}\gamma^{\beta}\gamma^{\gamma}\gamma_{\mu}
= - 2 \gamma^{\gamma} \gamma^{\beta} \gamma^{\alpha}$

\end{enumerate}

\begin{exercise}
  Calculate traces in $\overline{|\calM |^2}$:
\begin{eqnarray*}
\Tr[(\slash p_1 +m_1)\gamma^{\mu} (\slash p_2 -m_2)\gamma^{\nu}] &=& \mbox{?} \\
\Tr[(\slash p_4 -m_4)\gamma_{\mu} (\slash p_3 +m_3)\gamma_{\nu}] &=& \mbox{?}
\end{eqnarray*}
\end{exercise}

\begin{exercise}
  Calculate $\overline{|\calM |^2}$
\end{exercise}

\subsection{Kinematics in the center-of-mass frame}

In $e^+$$e^-$ coliders often $p_i \gg m_e, m_\mu$, $i=1,\ldots,4$, so
we can take

\begin{displaymath}
 m_i \to 0 \qquad \mbox{``high-energy'' or ``extreme relativistic'' limit}
\end{displaymath}

Then

\begin{equation}
\overline{|\calM |^2} = \frac{8e^4}{(p_1 + p_2)^4}
[(p_1 \cdot p_3)(p_2 \cdot p_4) + (p_1 \cdot p_4)(p_2 \cdot p_3)]
\end{equation}

To calculate scattering cross-section $\sigma$ we have to specialize
to some particular frame ($\sigma$ is \emph{not} frame-independent).
For $e^+$$e^-$ colliders the most relevant is the center-of-mass (CM) frame:

\vspace*{1ex}
\centerline{\includegraphics[scale=1.0]{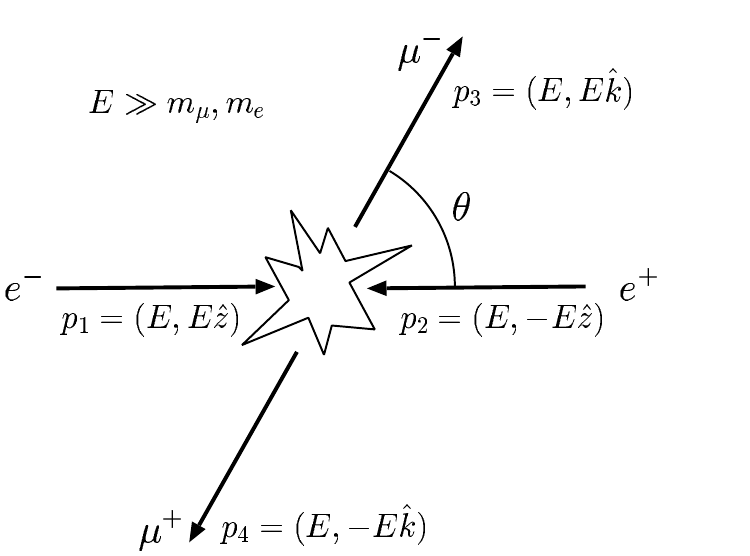}} 
\vspace*{1ex}

\begin{exercise}
 Express $\overline{|\calM |^2}$ in terms of $E$ and $\theta$.
\end{exercise}

\subsection{Integration over two-particle phase space}

Now we can use the ``golden rule'' (\ref{goldensigma}) for the
$1+2 \to 3+4$ differential  scattering cross-section

\begin{equation}
d\sigma = \frac{1}{u_\alpha}\frac{1}{2 E_1}\frac{1}{2 E_2}
 \overline{|{\cal M}|^2}\: d\mbox{Lips}_2
\end{equation}

where two-particle phase space to be integrated over is

\begin{equation}
d\mbox{Lips}_2=
 (2\pi)^4 \delta^4(p_1+p_2-p_3-p_4)
   \frac{d^3 p_3}{(2\pi)^3 \,2E_{3}} 
   \frac{d^3 p_4}{(2\pi)^3 \,2E_{4}}\;.
\end{equation}

 First we integrate over four out of six integration variables, and we
do this in general frame. $\delta$-function makes the integration over
$d^3 p_4$ trivial giving

\begin{equation}
 d\mbox{Lips}_2= \frac{1}{(2\pi)^2 \,4E_{3}E_{4}}\,
\delta(E_1 + E_2 - E_3 - E_4) \!\!\!\!\!\!\!
 \underbrace{d^3 p_3}_{\displaystyle \vec{p}_{3}^2
d|\vec{p}_3| d\Omega_3}
\end{equation}

Now we integrate over $d|\vec{p}_3|$ by noting that $E_3$ and $E_4$ are
functions of $|\vec{p}_3|$ 
\begin{eqnarray*}
E_3 & = & E_3(|\vec{p}_3|) = \sqrt{\vec{p}_{3}^2 + m_{3}^2}\;, \\
E_4 & = & \sqrt{\vec{p}_{4}^2 + m_{4}^2} = \sqrt{\vec{p}_{3}^2 + m_{4}^2} \;,
\end{eqnarray*}
 and by $\delta$-function relation
\begin{equation}
\delta(E_1 + E_2 - \sqrt{\vec{p}_{3}^2 + m_{3}^2} - 
\sqrt{\vec{p}_{3}^2 + m_{4}^2}) = \delta[f(|\vec{p}_3|)] =
\frac{\delta(|\vec{p}_3| - |\vec{p}^{(0)}_3| )}
{|f'(|\vec{p}_3|)|_{|\vec{p}_3|=|\vec{p}^{(0)}_3|}} \;.
\end{equation}
Here $|\vec{p}_3|$ is just the integration variable and $|\vec{p}^{(0)}_3|$
 is the
zero of $f(|\vec{p}_3|)$ i.e. the \emph{actual} momentum of the third particle.
After we integrate over $d|\vec{p}_3|$ we put 
$|\vec{p}^{(0)}_3| \to |\vec{p}_3|$.

Since
\begin{equation}
f'(|\vec{p}_3|) = - \frac{E_3 +E_4}{E_3 E_4} |\vec{p}_3| \;,
\end{equation}
we get
\begin{equation}
d\textrm{Lips}_2 = \frac{|\vec{p}_3| d\Omega}{16 \pi^2 (E_1 +E_2)}\;.
\end{equation}

Now we again specialize to the CM frame and note that the flux factor is
\begin{equation}
4 E_1 E_2 u_\alpha = 4\sqrt{(p_1 \cdot p_2)^2 - m_{1}^2 m_{2}^2}
= 4 |\vec{p}_1| (E_1 +E_2)\;,
\end{equation}
giving finally
\begin{equation}
\frac{d\sigma_{\mbox{\tiny CM}}}{d\Omega} = \frac{1}{64 \pi^2 (E_1 + E_2)^2}
\frac{|\vec{p}_3|}{|\vec{p}_1|} \overline{|\calM |^2}\;.
\end{equation}
Note that we kept masses in each step so this formula is generally valid
for any CM scattering.

 For our particular $e^- e^+ \to \mu^- \mu^+$ scattering this gives the
final result for differential cross-section (introducing the fine
structure constant $\alpha=e^2/(4\pi)$)
\begin{equation}
\frac{d\sigma}{d\Omega} = \frac{\alpha^2}{16 E^2}(1+\cos^2\theta)\;.
\end{equation}

\begin{exercise}
Integrate this to get the total cross section $\sigma$.
\end{exercise}

Note that it is obvious that $\sigma \propto \alpha^2$, and that
dimensional analysis requires  $\sigma \propto 1/E^2$, so only
angular dependence $(1+\cos^2\theta)$
tests QED as a theory of leptons and photons.

\subsection{Summary of steps}

  To recapitulate, calculating (unpolarized) scattering cross-section (or decay width)
consists of the following steps:

\begin{enumerate}
\item drawing the Feynman diagram(s)
\item writing $-i\calM$ using the Feynman rules
\item squaring $\calM$ and using the Casimir trick to get traces
\item evaluating traces
\item applying kinematics of the chosen frame
\item integrating over the phase space
\end{enumerate}

\subsection{Mandelstam variables}

Mandelstam variables $s$, $t$ and $u$ are often used in scattering
calculations. 
They are defined (for $1+2\to 3+4$ scattering) as
\begin{eqnarray*}
       s & = & (p_1 + p_2)^2 \\
       t & = & (p_1 - p_3)^2 \\
       u & = & (p_1 - p_4)^2 
\end{eqnarray*}

\begin{exercise}
Prove that $s+t+u = m_{1}^2 + m_{2}^2 + m_{3}^2 + m_{4}^2 $
\end{exercise}

This means that only two Mandelstam variables are independent. Their
main advantage is that they are Lorentz invariant which renders them
convenient for Feynman amplitude calculations. Only at the end
we can exchange them for ``experimenter's'' variables $E$ and $\theta$.

\begin{exercise}
Express $\overline{|\calM |^2}$ for $e^- e^+ \to \mu^- \mu^+$ scattering
in terms of Mandelstam variables.
\end{exercise}

\section*{Appendix: Doing Feynman diagrams on a computer}
\addcontentsline{toc}{section}{Appendix: Doing Feynman diagrams on a computer}

There are several computer programs that can perform some or all of the
steps in the calculation of Feynman diagrams. Here is a simple session
with one such program, \texttt{FeynCalc} \cite{Shtabovenko:2016sxi}
package for Wolfram's \emph{Mathematica},
where we calculate the same process, $e^- e^+ \to \mu^- \mu^+$, that we just
calculated in the text. 
Alternative framework, relying only on open source software is
FORM \cite{Vermaseren:2000nd}.

\centerline{\includegraphics[scale=0.75]{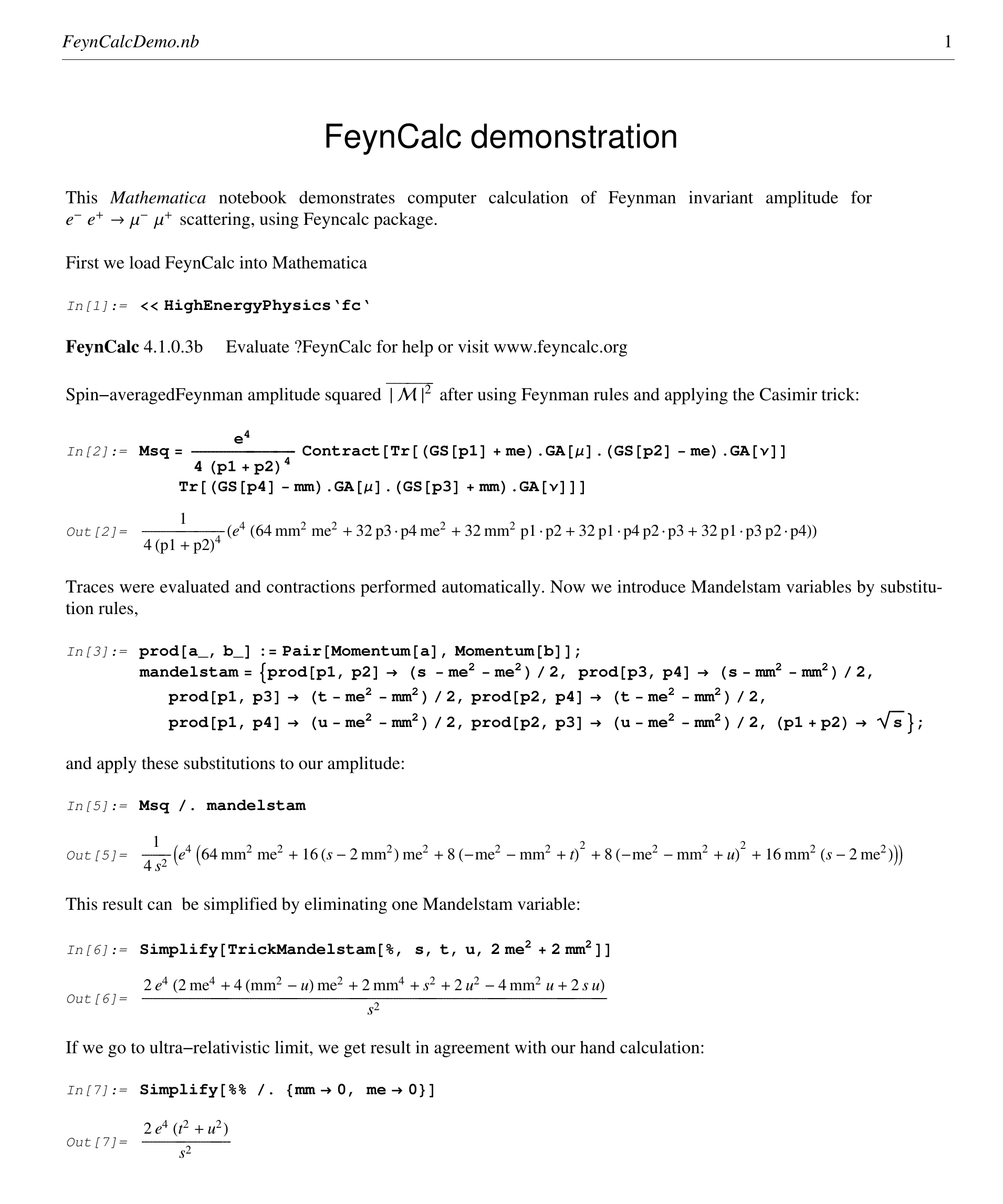}}

\end{document}